\documentstyle[12pt]{article}

\begin{document}

\begin{flushright}
IMSc/2003/04/05 \\
hep-th/0304152
\end{flushright} 

\vspace{2mm}

\vspace{2ex}

\begin{center}
{\large \bf Asymptotic Density of Open p-brane} \\ 

\vspace{2ex}

{\large \bf States with Zero-modes included } \\

\vspace{8ex}

{\large  S. Kalyana Rama}

\vspace{3ex}

Institute of Mathematical Sciences, C. I. T. Campus, 

Taramani, CHENNAI 600 113, India. 

\vspace{1ex}

email: krama@imsc.res.in \\ 

\end{center}

\vspace{6ex}

\centerline{ABSTRACT}
\begin{quote} 
We obtain the asymptotic density of open $p$-brane states with
zero-modes included. The resulting logarithmic correction to the
$p$-brane entropy has a coefficient $- \frac{p + 2}{2 p}$, and
is independent of the dimension of the embedding spacetime. Such
logarithmic corrections to the entropy, with precisely this
coefficient, appear in two other contexts also: a gas of
massless particles in $p$-dimensional space, and a Schwarzschild
black hole in $(p + 2)$-dimensional anti de Sitter spacetime.
\end{quote}

\vspace{2ex}

%PACS numbers: 11.25.-w

\newpage

\vspace{4ex}

{\bf 1.}  
The asymptotic density of states $\rho(N)$ at level $N$, 
$N \gg 1$, for $p$-branes compactified on $(S^1)^p \times 
{\bf R}^{D - p}$ has been be calculated within the semiclassical
quantisation scheme \cite{jackiw,others,gsw,b}. (For various
applications of this result see the review article in \cite{b}
and \cite{others,odin}.) The corresponding $p$-brane entropy,
given by ${\rm ln} \rho(N)$, then has a logarithmic correction
with a particular coefficient $X$, which depends on the
dimension $D$ of the embedding spacetime.

The correct counting of the asymptotic density of states must
also include the zero-mode states. They have been included for
the open string case ($p = 1$) in \cite{carlip,kaul}. As a
consequence, the logarithmic correction coefficient $X$ becomes
independent of $D$ and is given by $X = - \frac{3}{2}$. Such
logarithmic corrections to the entropy, with precisely this
coefficient, $- \frac{3}{2}$, have appeared in other contexts
also: in $(1 + 1)$-dimensional conformal field theories
\cite{carlip,kaul}, and in the entropies for $(2 + 1)$ and 
$(3 + 1)$ dimensional black holes calculated using the spin
network formalism \cite{km,sen}.

In this paper, using the results of \cite{b}, we obtain the
asymptotic density of states for open $p$-branes. We then
include the zero-modes following the methods of
\cite{carlip,kaul}. We find that the logarithmic correction
coefficient $X$ becomes independent of the dimension $D$ of the
embedding spacetime, and is given by $X = - \frac{p + 2}{2 p}$.

Logarithmic corrections to entropy also arise for statistical
mechanical systems due to statistical fluctuations \cite{dmb}.
Using the results of \cite{dmb}, we find that logarithmic
corrections to the entropy, with precisely the same coefficient
as that obtained in the open $p$-brane case, namely 
$- \frac{p + 2} {2 p}$, appear in two other contexts also: a gas
of massless particles in $p$-dimensional space, and a
Schwarzschild black hole in $(p + 2)$-dimensional anti de Sitter
spacetime \cite{dmb}.

This paper is organised as follows. In section {\bf 2}, we
briefly present the results of \cite{b} and, using them, obtain
the asymptotic density of states for open $p$-branes. In section
{\bf 3}, we include the zero-modes. In section {\bf 4}, we show
that such logarithmic corrections with precisely the same
coeffiecient appear in other contexts also. In section {\bf 5},
we conclude by mentioning a few issues for further study.

{\bf 2.}  
The asymptotic density of states $\rho(N)$ at level $N$, 
$N \gg 1$, for $p$-branes compactified on $(S^1)^p \times 
{\bf R}^{D - p}$ can be calculated within the semiclassical
quantisation scheme, and is of the form 
\begin{equation}\label{rho}
\rho(N) \simeq C N^B e^{A N^\delta}
\end{equation}
where $\delta$, $A$, $B$, and $C$ are constants. $\delta$ was
obtained in \cite{jackiw,others} and the correct expressions for
the remaining constants $A$, $B$, and $C$ in \cite{b}. The
corresponding $p$-brane entropy $S(N)$ is given by 
\begin{equation}\label{s}
S(N) = {\rm ln} \rho(N) 
\simeq S_0 + X \; {\rm ln} S_0 + (const) 
\end{equation}
where the leading term $S_0 = A N^\delta$ and 
$X = \frac{B}{\delta}$ is the coefficient of the logarithmic
correction to the entropy.

We now present briefly the results of \cite{b}. See \cite{b}
for details. In the semiclassical quantisation, the total
number operator ${\cal N}$ can be written in the proper time
formalism as
\begin{equation}\label{caln}
{\cal N} = \sum_{i = 1}^d \sum_{{\bf n} \ne {\bf 0}} 
\omega_{{\bf n}} {\cal N}_{{\bf n}}^i \; , 
\; \; \; \; 
\omega_{{\bf n}} \equiv \sqrt{\sum_{j = 1}^p n_j^2} \; ,                       
\end{equation} 
where $d = (D - p - 1)$, 
${\bf n} = (n_1, n_2, \cdots, n_p) \in {\bf Z}^p$,
${\bf 0} = (0, 0, \cdots, 0)$, and ${\cal N}_{{\bf n}}^i$ are
number operators \cite{duff}. For the sake of simplicity, we
have set the $p$-brane tension to unity and taken all the
circles in $(S^1)^p$ to be of unit radius. 

Let $\rho(N)$ be the number of independent eigenstates of the
total number operator ${\cal N}$ with eigenvalue $N$. Its
generating function $F(z)$ is given by 
\begin{equation}\label{f}
F(z) = \sum_{N = 0}^\infty \rho(N) e^{- z N} 
= Tr e^{- z N} = \prod_{{\bf n} \ne 0} 
\left( 1 - e^{- z \omega_{{\bf n}}} \right)^{- d} \; . 
\end{equation}
Inverting the above relation then gives $\rho(N)$ in terms of
$F(z)$: 
\begin{equation}\label{zcontour}
\rho(N) = - \frac{1}{2 \pi i} \oint d z \; 
e^{N z} \; F(z) 
\end{equation}
where the integration contour is a small circle around the
origin. Using Meinardus theorem \cite{meinardus} and the
properties of Epstein zeta function, an asymptotic expression
for $F(z)$ can be obtained in the limit $Re(z) \to 0$ which is
sufficient to obtain $\rho(N)$ in the limit $N \gg 1$. The
asymptotic expression for $F(z)$ is of the form
\cite{b}
\begin{equation}\label{fz}
F(z) \simeq c \;  z^b \; e^{a z^{- p}}
\end{equation}
where $a$, $b$, and $c$ are constants. The contour integral in
(\ref{zcontour}), and thus $\rho(N)$, can then be evaluated by
saddle point method. For $F(z)$ of the form given in equation
(\ref{fz}), the saddle point is located at $z = z_0 = \left(
\frac{N}{a p} \right)^{- \frac{1}{p + 1}}$ and $\rho(N)$, in the
limit $N \gg 1$, is given by
\begin{equation}\label{rhoN}
\rho(N) \simeq \frac{c \; (a p)^{\frac{2 b + 1}{2 (p + 1)}}}
{\sqrt{2 \pi (p + 1)}} \; 
N^{- \frac{2 b + p + 2}{2 (p + 1)}} \; 
e^{A N^{\frac{p}{p + 1}}} \; , \; \; \; 
A = \frac{p + 1}{p} \; (a p)^{\frac{1}{p + 1}} \; . 
\end{equation}
In our case, $a = \frac{2 d \; \Gamma(p) \pi^{\frac{p}{2}}
\zeta(p + 1)} {\Gamma(\frac{p}{2})}$ and $b = d$ where 
$d = (D - p - 1)$ and $\zeta$ is the Riemann zeta function; the
constant $c$ is given explicitly in \cite{b} and is not required
for our purposes.  See \cite{b} for further details.

We now obtain the asymptotic density of states for an open
$p$-brane using the above results. In this case modes ${\bf n}$
and ${\bf - n}$ together contribute to one standing wave (SW)
mode of the open $p$-brane and, hence, should be counted only
once. Thus, the corresponding total number operator is given by 
\begin{equation}\label{calno}
{\cal N} = \sum_{i = 1}^d \; \sum_{{\rm SW}; {\bf n} 
\ne {\bf 0}} \omega_{{\bf n}} {\cal N}_{{\bf n}}^i \; . 
\end{equation} 
The corresponding generating function $F_o(z)$ is given by 
\begin{equation}\label{fo}
F_o(z) = \sum_{N = 0}^\infty \rho_o(N) e^{- z N} 
= \prod_{{\rm SW}; \; {\bf n} \ne 0} 
\left( 1 - e^{- z \omega_{{\bf n}}} \right)^{- d}  
= \prod_{{\bf n} \ne 0} \left( 
1 - e^{- z \omega_{{\bf n}}} \right)^{- \frac{d}{2}} 
\end{equation} 
where the last equality follows since $\omega_{{\bf n}} =
\omega_{{\bf - n}}$ and the product in the last expression
includes both ${\bf n}$ and ${\bf - n}$. Note that the
generating function $F_o(z)$ is identical to that in equation
(\ref{f}) with $d$ there replaced by $\frac{d}{2}$. It therefore
follows that the density of states $\rho_o(N)$ for an open
$p$-brane in the limit $N \gg 1$ is given by (\ref{rhoN}), but
now with $d$ replaced by $\frac{d}{2}$. Explicitly, in the limit
$N \gg 1$, 
\begin{equation}\label{rhoo}
\rho_o(N) \simeq \frac{c \; (a p)^{\frac{D - p}{2 (p + 1)}}}
{\sqrt{2 \pi (p + 1)}} \; 
N^{- \frac{D + 1}{2 (p + 1)}} \; 
e^{A N^{\frac{p}{p + 1}}} \; , \; \; \; 
A = \frac{p + 1}{p} \; (a p)^{\frac{1}{p + 1}} 
\end{equation}
where $a = \frac{(D - p - 1) \; \Gamma(p) \pi^{\frac{p}{2}}
\zeta(p + 1)} {\Gamma(\frac{p}{2})}$, $c$ is as given in
\cite{b} but with $d = (D - p - 1)$ there replaced by
$\frac{d}{2} = \frac{D - p - 1}{2}$, and we have used $b =
\frac{d}{2} = \frac{D - p - 1}{2}$ in obtaining (\ref{rhoo}).
Note that $\rho(N)$ is of the form given in equation (\ref{rho})
with $\delta = \frac{p} {p + 1}$ and $B = - \frac{D + 1}
{2 (p + 1)}$, and that string theory result \cite{gsw} is
obtained upon setting $p = 1$.

{\bf 3.}
We now include the zero-modes and obtain the resulting
asymptotic density of open $p$-brane states. The complete
Hamiltonian for a $p$-brane in the proper time formalism is
given by 
\begin{equation}\label{h}
H = {\bf p}^2 + {\cal N}
\end{equation}
where ${\bf p}^2$ is the transverse momentum square opeartor and
${\cal N}$ is the total number operator given in (\ref{calno}),
and they both commute with each other \cite{b}.

The correct counting of the total number of states $\rho(N)$
must include the zero-mode states also, namely those
corresponding to the transverse momentum. For open strings 
($p = 1$), $\rho(N)$ with zero-modes included has been
calculated in \cite{carlip,kaul} in the limit $N \gg 1$. The
resulting $\rho(N)$ is still given by equation (\ref{rho}) with
$\delta = \frac{1}{2}$ as before, but now with 
$B = - \frac{3}{4}$ independent of the dimension $D$ of the
embedding spacetime. See \cite{carlip,kaul} for further
discussions.

The total number of states $\rho(N)$ for open $p$-branes, with
zero-modes included, can be calculated for other values of $p$
also in the limit $N \gg 1$. $\rho(N)$ is still given by
equation (\ref{zcontour}), but now the relevent generating
function $F(z)$, with zero-modes included, is given by
\begin{equation}\label{fcorrect}
F(z) = Tr e^{- z H} = \left( \int d^{D - p- 1}p \; 
e^{- z {\bf p}^2} \right) \; Tr e^{- z N} 
= \left( \int d^{D - p- 1}p \; 
e^{- z {\bf p}^2} \right) \; F_o(z) 
\end{equation}
where the second equality follows since the operators 
${\bf p}^2$ and ${\cal N}$ commute with each other and $F_o(z)$
is the generating function given in equation (\ref{fo}). The
momentum integral can be evaluated easily and results in an
extra $z$-dependent factor given by 
\[
\int d^{D - p- 1}p \; e^{- z {\bf p}^2} 
= c_0 \; z^{- \frac{D - p - 1}{2}} 
\] 
where $c_0$ is a constant. The contour integral in
(\ref{zcontour}), and thus $\rho(N)$, can now be evaluated by
saddle point method. The saddle point is at $z = z_0 = \left(
\frac{N} {a p} \right)^{- \frac{1} {p + 1}}$ as before, and the
total number of open $p$-brane states $\rho_o(N)$, in the limit
$N \gg 1$ and with zero-modes included, is now given by
\begin{equation}\label{rhoN0}
\rho_o(N) \simeq \frac{c c_0 \; (a p)^{\frac{1}{2 (p + 1)}}}
{\sqrt{2 \pi (p + 1)}} \; N^{- \frac{p + 2}{2 (p + 1)}} 
\; e^{A N^{\frac{p}{p + 1}}}  
\end{equation}
where the constants $a$ and $c$ are as in equation
(\ref{rhoo}). Clearly, $\rho(N)$ is of the form given in
equation (\ref{rho}) with $\delta = \frac{p} {p + 1}$ as before,
but now with $B = - \frac{p + 2} {2 (p + 1)}$ independent of the
dimension $D$ of the embedding spacetime.

The open $p$-brane entropy $S(N)$ with zero-modes included is
thus given by 
\begin{equation}\label{sp}
S(N) = {\rm ln} \rho(N) \simeq S_0 - \frac{p + 2}{2 p} \; 
{\rm ln} S_0 + (const) 
\end{equation} 
where $S_0 = A N^{\frac{p}{p + 1}}$. The coefficient of the
logarithmic correction to the open $p$-brane entropy with
zero-modes included is now given by $- \frac{p + 2}{2 p}$ and is
independent of the dimension $D$ of the embedding spacetime.
Note that without zero-modes included it is given by 
$- \frac{D + 1}{2 p}$, as can be seen from equations (\ref{s})
and (\ref{rhoo}), and depends on $D$.

{\bf 4.}  
For open strings ($p = 1$) with zero-modes included, the
coefficient of the logarithmic correction to the entropy given
above becomes $- \frac{3}{2}$, which agrees with the results of
\cite{carlip,kaul}. Such logarithmic corrections, with precisely
this coefficient, $- \frac{3}{2}$, have appeared in other
contexts also: in $(1 + 1)$-dimensional conformal field theories
\cite{carlip,kaul}, and in the entropies for $(2 + 1)$ and 
$(3 + 1)$ dimensional black holes calculated using the spin
network formalism \cite{km,sen}.

Similarly, it turns out that logarithmic corrections, with
precisely the coefficient given in equation (\ref{sp}), namely
$- \frac{p + 2}{2 p}$, also appear in two other contexts.
Recently, logarithmic corrections to entropies of statistical
mechanical systems, arising due to statistical fluctuations,
have been obtained in \cite{dmb}. One calculates the density of
states $\rho(E)$ as an inverse Laplace transformation of the
partition function in the canonical ensemble. Statistical
fluctuations can then be incorporated naturally. Then, 
$\rho(E) \Delta$ is the number of states with energy in the
range $E \pm \frac{\Delta}{2}$ where $\Delta$ depends on the
precision with which the system is prepared and, in particular,
is independent of $E$. The entropy $S(E)$ is therefore given by
$S(E) = {\rm ln} \rho(E) + (const)$.

The result of \cite{dmb} is that for a system at temperature
$T$, with specific heat ${\cal C}$ (which must be positive for
this formalism to be applicable), one obtains for its entropy 
\begin{equation}\label{dmb}
S = S_0 - \frac{1}{2} {\rm ln} ({\cal C} \; T^2)+ (const)
\end{equation}
where $S_0$ is the leading term. See \cite{dmb} for details.

Now, consider a gas of massless particles in $p$-dimensional
space. Then 
\[
S_0 \propto T^p \; , \; \; \; 
E \propto T^{p + 1} \; , \; \; \; 
{\cal C} \propto T^p \; . 
\]
Equation (\ref{dmb}), therefore, gives 
\begin{equation}\label{gas}
S = S_0 - \frac{p + 2}{2 p} {\rm ln} S_0 + (const) \; . 
\end{equation}

For a Schwarzschild black hole, the specific heat is negative
and, hence, the above formalism is inapplicable \cite{dmb}.
However, a Schwarzschild black hole of sufficiently large mass
in a $d$-dimensional anti de Sitter spacetime (AdS$_d$) has
positive specific heat. Consider an AdS$_{p + 2}$ Schwarzschild
black hole of mass $M$. For sufficiently large $M$, one has
\cite{adswitten} 
\[
S_0 \propto r_+^p \; , \; \; \; 
E = M \propto r_+^{p + 1} \; , \; \; \; 
T \propto r_+ \; , 
\]
where $r_+$ is the horizon. It then follows that 
${\cal C} \propto r_+^p$. Equation (\ref{dmb}), therefore, gives
\begin{equation}\label{ads}
S = S_0 - \frac{p + 2}{2 p} {\rm ln} S_0 + (const) \; . 
\end{equation}
See \cite{dmb}) for details. For AdS$_3$, see also
\cite{km,sen}. From equations (\ref{gas}) and (\ref{ads}),
we see that logarithmic corrections to the entropy of a gas of
massless particles in $p$-dimensional space, and to that of an
AdS$_{p + 2}$ Schwarzschild black hole, both have a coefficient
$- \frac{p + 2} {2 p}$ which is precisely the same as that
obtained in the open $p$-brane case with zero-modes included.

{\bf 5.}   
To summarise, we have obtained the asymptotic density of open
$p$-brane states with zero-modes included. The corresponding
open $p$-brane entropy has a logarithmic correction, with a
coefficient $- \frac{p + 2}{2 p}$. Such logarithmic corrections,
with precisely the same coefficient, also appear for a
$p$-dimensional gas and for an AdS$_{p + 2}$ Schwarzschild black
hole where the corrections arise due to statistical
fluctuations.

The relation of a $p$-dimensional gas to AdS$_{p + 2}$
Schwarzschild black hole, for $p = 1, 2, 3$, and $5$, can be
understood in the context of AdS/CFT duality \cite{adscft} as
that of a boundary conformal field theory at high temperature
\cite{adswitten}. In light of the present results, one may
explore the relations between quantum/semi classical $p$-branes,
$p$-dimensional gas, and AdS$_{p + 2}$ spacetime in more
detail. In particular, it will be interesting to know if the
values of $p$ are restricted for quantum/semi classical
$p$-branes, or if a duality exists between $p$-dimensional gas
and AdS$_{p + 2}$ spacetime for any value of $p$.

As mentioned earlier, the coefficient $X = - \frac{3}{2}$,
corresponding to $p = 1$, also appears for the entropy of a 
$(3 + 1)$ dimensional black hole calculated using the spin
network formalism \cite{km}.  In this formalism, one considers
punctures, each carrying a spin ${\bf J}_{puncture}$, and counts
the number of spin singlet states, namely those states with
${\bf J}(total) = 0$. The leading term (and a part of the
logarithmic correction) in the entropy corresponds to the number
of states with $J_z(total) = 0$. However, such states include
states with ${\bf J}(total) \ne 0$ also. A correct counting,
that counts states with ${\bf J}(total) = 0$ only, then leads to
the coefficient $- \frac{3}{2}$ for the logarithmic correction
to the entropy \cite{km}. It will be interesting to find if a
similar interpretation exists for the logarithmic correction
coefficient $X = - \frac{p + 2} {2 p}$ for other values of $p$
also.

\vspace{3ex}

{\bf Acknowledgement:} 
We thank C. Castro for suggesting this problem, pointing out the
review article in \cite{b}, and for discussions; R. Kaul for
discussions regarding reference \cite{kaul}; and K. Kirsten for
correspondence.

%\newpage

\end{document}